\documentclass[letterpaper, 10 pt, conference]{ieeeconf}  

\IEEEoverridecommandlockouts                              
\usepackage{float}
\usepackage{amsfonts,mathrsfs,bbm,amsmath,stmaryrd,amssymb,pifont}%
\usepackage{amscd,graphicx,array,dsfont,latexsym,xcolor}%
\usepackage{cite}

\newcommand{\qed}{\tag*{$\Box$}}

\newtheorem{assumption}{Assumption}
\newtheorem{lemma}{Lemma}
\newtheorem{theorem}{Theorem}
\newtheorem{remark}{Remark}

\newtheorem{definition}{Definition}

\newtheorem{proposition}{Proposition}

\newcommand{\R}{\mathbb{R}}

\usepackage[prependcaption,colorinlistoftodos]{todonotes}

\title{\LARGE \bf
Learning control for polynomial systems using sum of squares relaxations
}

\author{Meichen Guo, Claudio De Persis, and Pietro Tesi
\thanks{Meichen Guo and Claudio De Persis are with ENTEG, Faculty of Science and Engineering, University of Groningen, 9747 AG Groningen, The Netherlands. Email: {\tt\small meichen.guo@rug.nl, c.de.persis@rug.nl} Pietro Tesi is with DINFO, University of Florence, 50139 Florence, Italy. E-mail: {\tt\small pietro.tesi@unifi.it}}
}

\begin{document}

\maketitle
\thispagestyle{empty}
\pagestyle{empty}

\begin{abstract}
This paper considers the problem of learning
control laws for nonlinear polynomial systems directly from the data, which are input-output measurements collected in an experiment over a finite time period. Without explicitly
identifying the system dynamics, stabilizing laws are directly
designed for nonlinear polynomial systems using experimental data alone. By using data-based sum of square programming, the stabilizing state-dependent control gains
can be constructed.
\end{abstract}

\section{Introduction}
Natural and engineering systems often have nonlinear dynamics that are important to the scientific understanding of those systems. However, analysis and control of nonlinear systems has always been challenging. Learning based control has drawn much attention in the control community over the past decades, where properties of the unknown dynamics can be learned from data. When the explicit model of the controlled system is unknown, a model is often first identified from input-output measurements of the system, and then a model-based controller can be designed. Although there are many well-known and widely-used methods for linear system identification, current nonlinear system identification methods still have many limitations \cite{Kerschen2006identification}. Moreover, the identification of nonlinear dynamics can be extremely difficult and time consuming. These limitations of current nonlinear learning control methods have motivated us to develop an approach that learns control laws directly from the input-output data, without explicitly identifying the nonlinear system model.

Without explicit model identification, various learning approaches have been used to control nonlinear systems. For example, virtual reference feedback tuning (VRFT) selects the controller via an off-line model reference optimization performed based on the data \cite{campi2006direct}. The authors of \cite{Fliess2009} approximate the nonlinear dynamics as a linear model in a sufficiently short time interval and then design an intelligent PID controller for the model. Following the philosophy of \cite{Fliess2009}, the authors of \cite{Tabuada2017CDC} prove the existence of a sufficiently high sampling rate with which the controller designed for the approximated linear model guarantees the control performance of the true nonlinear dynamics. Another method for nonlinear learning control involves adaptive dynamic programming, such as in \cite{Lee2005nonlinearQlearning}, where on-line closed-loop experiment and off-line controller redesign are performed in each iteration step. Despite these works on nonlinear learning control, the open problem remains how to learn a control law directly from the input-output data without model approximation or iterative experiments and redesigns.

For linear discrete-time systems, \cite{DDC2019} has shown that feedback control laws can be directly learned from data using Willems \emph{et al.}'s fundamental lemma and writing a data-dependent representation of the system. Particularly, \cite{DDC2019} parameterized the feedback controller using the input-output data and transforms the design of a feedback gain into solving a data-dependent linear matrix inequality (LMI). Furthermore, by representing the unknown nonlinear dynamics as the sum of a linear model and a noise term containing the higher-order terms of the nonlinearity, this approach can also be used to stabilize an unstable equilibrium of a nonlinear discrete-time system. The results of \cite{DDC2019} has led us to the questions whether Willems \emph{et al.}'s fundamental lemma can be used for learning control of other classes of nonlinear systems without system identification, and whether a computationally tractable approach can be developed to find a stabilizing nonlinear control law.

To answer these questions, this paper considers learning stabilizing control laws for a class of continuous time linear-like polynomial systems directly from data. By collecting input-output measurements in an open loop experiment of the unknown dynamics and arranging them in the form of Hankel matrices, a data-based representation can be written for the closed-loop system. For a special class of Lyapunov candidates, learning of a stabilizer requires constructing nonnegative polynomial matrices. This kind of problems are often computationally intractable, in the sense that the problem cannot be practically solved by any algorithm within reasonable time. One of the computationally tractable approaches for model-based nonlinear control is the sum of squares (SOS) optimization, which can be solved through semi-definite programming (SPD) as shown in \cite{Prajna2004SOS}. If a polynomial can be decomposed as an SOS, then it is globally positive semi-definite \cite{Choi1995sos}. This argument used jointly with SPD can relax the computation for proving global positive semi-definiteness of multivariate polynomials \cite{ParriloThesis}. Specifically, a Lyapunov-based control synthesis is proposed in \cite{Prajna2004SOS} for a class of nonlinear systems in a linear-like state-dependent form, such that the design of the feedback controller can mimic the pure linear ones. By choosing a special class of Lyapunov candidates, a stabilizing feedback control gain can be found by solving state-dependent SOS programs.

In this work, we formulate the data-driven stabilizer design for polynomial systems into SOS programs depending on experimental data alone. The highlights of the \emph{learning} or
\emph{data-driven} control approach proposed in this paper include (i) the learning of nonlinear control law is free of model identification; (ii) finite data collected in one experiment is needed for stabilizing the system; (iii) the computation of the control gain is tractable.

The rest of the paper is arranged as follows. Preliminaries on SOS matrix polynomials, model-based stabilization of polynomial systems, and data-driven control of linear systems are presented in Section~\ref{sec_preliminaries}. The data-dependent representation of the polynomial system and its data-driven stabilization are developed in Section~\ref{sec_datadrivendesign}. Section~\ref{sec_example} shows an example that is stabilized by the proposed SOS program. Finally, some conclusive remarks and ideas for future works are given in Section~\ref{sec_conclusion}.

\emph{Notations.} The following notations are adopted throughout the paper:
\begin{itemize}
  \item[--] $A\succeq 0$: matrix $A$ is positive semi-definite;
  \item[--] $A\succ 0$: matrix $A$ is positive definite;
  \item[--] $A\succeq B$: matrix $A-B$ is positive semi-definite;
  \item[--] $A\otimes B$: Kronecker product of matrices $A$ and $B$;
  \item[--] $\mathbb{N}$: the set of natural numbers including $0$;
  \item[--] $\mathbb{N}_{>0}$: the set of natural numbers excluding $0$;
  \item[--] $\R$: the set of real numbers;
  \item[--] $\R_{>0}$: the set of positive real numbers;
  \item[--] $\mathbb{S}^{n}$: the set of $n \times n$ symmetric matrices.
\end{itemize}

\section{Preliminaries}\label{sec_preliminaries}
In this section, some important notions of SOS polynomials, results on stabilization of polynomial systems, and preliminaries on data-driven control of linear discrete time systems are revisited.
\subsection{SOS matrix polynomials}
Following \cite{Prajna2004SOS,Chesi2009Book,Chesi2010}, we first present some important definitions and properties on SOS polynomials.

A function $h:\R^{n}\rightarrow\R$ is a \emph{monomial} of degree $d$ in $n$ scalar variables if
\begin{align}\label{definition_form}
  h(x)=a_{q}x^{q}
\end{align}
where $a_{q}\in\R$, $x\in\R^{n}$, $x^{q}=x_{1}^{q_{1}}~x_{2}^{q_{2}}~\dots~x_{n}^{q_{n}}$, and
\begin{align}
  q\in
  \Big{\{} q\in\mathbb{N}^{n}:~\sum^{n}_{i=1}q_{i}=d \Big{\}}.
\end{align}
A function $f$ is a polynomial if it is a sum of monomials $h_{1},h_{2},\dots$ with finite degree. Denote the set of polynomials as $\mathcal{P}$. The largest degree of $h_{i}$ is the degree of $f$. A function $M:\R^{n}\rightarrow\R^{r\times r}$ is a matrix polynomial if every entry of $M$ satisfies $M_{ij}\in\mathcal{P}$ for all $i,j=1,\dots,r$. The largest degree of $M_{ij}$ is the degree of $M$. The set of matrix polynomials of size $r$ is denoted by $\mathcal{P}_{r}$.

\smallskip

The matrix case of SOS polynomials is defined as follows.

\begin{definition}
(\emph{SOS matrix polynomial}\cite{Chesi2010})
$M\in\mathcal{P}_{r}$ is an SOS matrix polynomial if there exist $M_{1},\dots,M_{k}\in\mathcal{P}_{r}$ such that
\begin{align}\label{def_SOSmatrix}
  M(x)=\sum^{k}_{i=1}M_{i}(x)^{\top}M_{i}(x).
\end{align}
\end{definition}

\smallskip

Some important properties of SOS matrices are summarized in the following proposition.
\smallskip

\begin{proposition}\label{prop_sos}
  (\emph{Properties of SOS Matrix})
  For a matrix polynomial $M\in\mathcal{P}_{r}$, consider the following conditions
  \begin{enumerate}
    \item[(i)] $M(x)$ is SOS;
    \item[(ii)] $M(x)\succeq 0$ for all $x\in\R^{n}$;
    \item[(iii)] the polynomial $y^{\top}M(x)y$ is SOS in the extended variable $[x^{\top}~y^{\top}]^{\top}$, where $y\in\R^{r}$.
  \end{enumerate}
  Then, (i) $\Rightarrow$ (ii) and (i) $\Leftrightarrow$ (iii). 
\end{proposition}

\medskip

\proof
The proof of (i) $\Rightarrow$ (ii) comes directly from the definition of SOS matrix polynomial (\ref{def_SOSmatrix}). The equivalence of (i) and (iv) is given in \cite[Theorem 1.6]{Chesi2009Book}.
\hfill$\square$

\subsection{Model-based stabilization of polynomial systems}

Consider a class of nonlinear polynomial systems having the form
\begin{align}\label{dynamics_nonlinear poly}
\dot{x}=A(x)Z(x)+B(x)u,
\end{align}
where $A$ and $B$ are matrix polynomials in $x$ and $Z(x)$ is an $N\times 1$ vector of monomials in $x$. The following assumption on $Z(x)$ ensures that the origin is an equilibrium of (\ref{dynamics_nonlinear poly}).

\medskip

\begin{assumption}\label{assumption_equilibirium}
  Vector $Z(x)=0$ if and only if $x=0_{n\times 1}$.
\end{assumption}

\medskip

Based on \cite[Theorem 6]{Prajna2004SOS}, a result on nonlinear control of system (\ref{dynamics_nonlinear poly}) using SOS is presented as follows.

\medskip

\begin{proposition}\label{prop_stability}
  (\emph{Nonlinear stabilization using SOS})

  For the nonlinear polynomial system (\ref{dynamics_nonlinear poly}), under Assumption \ref{assumption_equilibirium}, if there exist a symmetric constant matrix $P$ and a matrix polynomial $Y(x)\in\R^{m\times N}$ such that
  \begin{itemize}
    \item[(i)] $P\succ 0$, and
    \item[(ii)] the matrix polynomial
   \begin{align*}
  Q(x)&:=-
  \frac{\partial Z}{\partial x}
  \begin{bmatrix}   A(x)  & B(x)   \end{bmatrix}
  \begin{bmatrix}  P \\ Y(x) \end{bmatrix} \\
  &\quad -\begin{bmatrix}  P \\ Y(x) \end{bmatrix}^{\top}   \begin{bmatrix}   A(x)  & B(x)   \end{bmatrix}^{\top}
  \Big( \frac{\partial Z}{\partial x}\Big)^{\top} -\epsilon(x)I_{N}
  \end{align*}
  is SOS for some SOS polynomial $\epsilon(x)$,
  \end{itemize}
  then the controller
  \begin{align*}
  u= F(x) Z(x):=Y(x)P^{-1}Z(x)
  \end{align*}
  stabilizes the polynomial system. Moreover, if $\epsilon(x)>0$ for all $x\ne 0$, the zero equilibrium is asymptotically stable.
\end{proposition}

\medskip

The proof of Proposition \ref{prop_stability} follows that of \cite[Theorem 6]{Prajna2004SOS} and the equivalence of (i) and (iii) in Proposition \ref{prop_sos}.

\medskip

\subsection{Data-driven stabilization of linear systems}

Using Willems \emph{et al.}'s fundamental lemma, \cite{DDC2019} gives a data-dependent representation of the closed-loop dynamics of discrete-time linear systems under feedback interconnection. Particularly, \cite{DDC2019} considers a controllable and observable discrete-time linear system
\begin{align}\label{dynamics_LDT}
  x(k+1) &= Ax(k)+Bu(k),
\end{align}
where $x\in\R^{n}$ and $u\in\R^{m}$. During an experiment over the time interval $[0,T-1]$, the input-output data collected are arranged in the form of Hankel matrix as
\begin{align*}
  U_{0,1,T}&:=\begin{bmatrix}
              u(0)\! & \!u(1)\! & \!\cdots\! &\! u(T-1)
            \end{bmatrix},\\
  X_{0,T}&:=\begin{bmatrix}
              x(0)\! & \!x(1)\! & \!\cdots\! &\! x(T-1)
            \end{bmatrix},\\
  X_{1,T}&:=\begin{bmatrix}
              x(1)\! & \!x(2)\! & \!\cdots\! &\! x(T)
            \end{bmatrix}.
\end{align*}
Under state feedback controller $u=Kx$, a data parametrization of (\ref{dynamics_LDT}) is given in the follow result from \cite{DDC2019}.

\medskip

\begin{proposition}
  Let the input-output data satisfy $\mathrm{rank}(X_{0,T})=n$, compute the matrix $G_{K}$ such that
  \begin{align}\label{definition_GK}
    I_{n}=X_{0,T}G_{K}
  \end{align}
  and set
  \begin{align*}
  K=U_{0,1,T}G_{K}.
  \end{align*}
  Then the closed-loop system of (\ref{dynamics_LDT}) with state feedback controller $u=Kx$ has the data-based representation
  \begin{align}
    x(k+1)=X_{1,T}G_{K}x(k).
  \end{align}
\end{proposition}

\medskip

Using this data-based representation of the closed-loop system, the control gain $K$ can be designed directly based on data without explicit identification of the system matrices $A$
and $B$. In fact, the closed-loop system dynamics under $u=Kx$ is
\begin{align}
  A+BK=X_{1,T}G_{K}
\end{align}
with $G_{K}$ as defined in (\ref{definition_GK}). Thus, the design of a stabilizing gain $K$ becomes searching for a matrix $G_{K}$ such that $X_{1,T}G_{K}$ satisfies the classic Lyapunov stability condition.

A similar representation also holds for linear continuous-time systems $\dot x = Ax +Bu$. In this case, the matrix $X_{1,T}$  contains the derivatives of the states at the sampling times when the measurements are taken (see \cite[Remark 2]{DDC2019}). In this paper, we will focus on continuous-time polynomial systems because this allows us to adopt the tools from \cite{Chesi2009Book, Chesi2010, Prajna2004SOS}, while for discrete-time polynomial systems less results are available \cite{tedrake2018iros,Pirkelmann2019nolcos}.
\medskip

\section{Data-Based Representation and Data-Driven Stabilization of Nonlinear Polynomials Systems}
\label{sec_datadrivendesign}
Inspired by the stabilization of nonlinear polynomial systems using SOS and the data-driven stabilization of linear systems, we aim to design data-driven controllers for a class of nonlinear polynomial systems that is linear in the vector of state monomials.

\subsection{Data-based system representation}

We consider a simplified version of (\ref{dynamics_nonlinear poly}) in the form of
\begin{align}\label{dynamics_non poly simple}
  \dot{x} = AZ(x)+Bu
\end{align}
where $Z(x)\in\R^{N}$ is a vector of monomials in state $x\in\R^{n}$, $u\in\R^{m}$ is the control input, and $A\in\R^{n\times n}$ and $B\in\R^{n\times m}$ are unknown constant matrices.

The SOS approach to the control of nonlinear polynomial systems (\ref{dynamics_non poly simple}) mimics the linear case and takes the vector of monomials $Z(x)$ as the counterpart of the state $x$. We assume that (\ref{dynamics_non poly simple}) has matrix $B$ independent of $x$ for the sake of simplicity. Similar to the linear case, the control input here is designed as
\begin{align}\label{controller}
u= F(x) Z(x)
\end{align}
where $F(x)$ is to be determined. As it is assumed that the system matrices $A$ and $B$ are unknown, the objective of this paper is to use data collected in an experiment to design the feedback gain $F(x)$ directly.

In this paper, we consider nonlinear polynomial systems with unknown model (\ref{dynamics_non poly simple}). In practice, some \emph{a priori} information, such as physical considerations, can be used to find insights on the most appropriate choice of $Z(x)$.

In an experiment over the time interval $[t_{0},t_{0}+(T-1)\tau]$ where $T\in\mathbb{N}_{>0}$ is the number of collected samples, and $\tau\in\R_{>0}$ is the sampling time, the Hankel matrices of the sampled input-output data are defined as
\begin{align*}
  U_{0,1,T}&:=\begin{bmatrix}
              u(t_{0})\! & \!u(t_{0}\!+\!\tau)\! & \!\cdots\! &\! u(t_{0}+(T\!-\!1)\tau)
            \end{bmatrix},\\
  X_{0,T}&:=\begin{bmatrix}
              x(t_{0})\! & \!x(t_{0}\!+\!\tau)\! & \!\cdots\! &\! x(t_{0}+(T\!-\!1)\tau)
            \end{bmatrix},\\
  X_{1,T}&:=\begin{bmatrix}
              \dot{x}(t_{0})\! & \!\dot{x}(t_{0}\!+\!\tau)\! & \!\cdots\! &\! \dot{x}(t_{0}+(T\!-\!1)\tau)
            \end{bmatrix}.
\end{align*}
Using the vector $Z(x)$ and the samples $X_{0,T}$,  we can calculate the matrix
\begin{align*}
  &\mathcal{Z}_{0,T}\!\\
  &:=\!\begin{bmatrix}
              Z(x(t_{0}))\! & \!Z(x(t_{0}\!+\!\tau))\! & \!\cdots\! &\! Z(x(t_{0}\!+\!(T\!-\!1)\tau))
            \end{bmatrix}\!.
\end{align*}

\medskip

\begin{assumption}\label{assumption_PE}
  The $N\times T$ matrix $\mathcal{Z}_{0,T}$ has full row rank.
\end{assumption}

\medskip

\begin{remark}(\emph{Full row rank of $\mathcal{Z}_{0,T}$})
For Assumption \ref{assumption_PE} to hold, the number of samples $T$ must satisfy $T\ge N$. Note that, since the matrix $\mathcal{Z}_{0,T}$ is computable from the data, this assumption is
verifiable. 
\end{remark}

\medskip

\begin{lemma}\label{lemma_data rep}
Under Assumption \ref{assumption_PE}, let $G(x)$ be a $T\times N$ matrix such that
\begin{align}\label{condition_Gx}
I_{N}=\mathcal{Z}_{0,T}G(x)
\end{align}
and set $F(x)=U_{0,1,T}G(x)$ in (\ref{controller}). Then system (\ref{dynamics_non poly simple})
in closed-loop with the state-feedback controller (\ref{controller}) has the following data-dependent equivalent representation
\begin{align}
\dot x = X_{1,T} G (x) Z(x).
\end{align}
\end{lemma}
\medskip

\proof Under Assumption \ref{assumption_PE}, a matrix $G(x)$ such that (\ref{condition_Gx}) holds exists. Note that
\begin{align}\label{expression_Gx}
G(x)=  \mathcal{Z}_{0,T} ^\dag+ (I_T-  \mathcal{Z}_{0,T}^\dag  \mathcal{Z}_{0,T})w(x),
\end{align}
where $\mathcal{Z}_{0,T}^\dag= \mathcal{Z}_{0,T}^\top \left(\mathcal{Z}_{0,T} \mathcal{Z}_{0,T}^\top\right)^{-1}$, $w(x)$ is a $T\times N$ matrix polynomial in $x$, and $(I_T-  \mathcal{Z}_{0,T}^\dag  \mathcal{Z}_{0,T})$ is the projection onto $\ker (\mathcal{Z}_{0,T})$. Since
\begin{align*}
F(x)= U_{0,1,T} G(x)
\end{align*}
the closed-loop system can be written as
\begin{align*}
(A + B F(x)) Z(x)&= \begin{bmatrix}  B &  A \end{bmatrix}
 \begin{bmatrix} F(x) \\ I_N \end{bmatrix} Z(x)\\
 &= \begin{bmatrix}  B & A \end{bmatrix}
\begin{bmatrix}
U_{0,1,T}\\
\mathcal{Z}_{0,T}
\end{bmatrix}
G (x) Z(x)
\end{align*}
By the dynamics of the system, it holds that
\begin{align*}
\begin{bmatrix} B & A \end{bmatrix}
\begin{bmatrix}
U_{0,1,T}\\
\mathcal{Z}_{0,T}
\end{bmatrix}= X_{1,T}.
\end{align*}
Therefore, we obtain that
\begin{align*}
  \dot x = ( A + B F(x)) Z(x)= X_{1,T} G (x) Z(x).
  \qed
\end{align*}

\subsection{Data-driven stabilization using SOS relaxations}
Using the data-based representation in Lemma \ref{lemma_data rep}, we can design the control gain $F(x)$ by searching for a $G(x)$ that satisfies some stabilizing criteria. Particularly, we find inspiration from Proposition \ref{prop_stability} and present the following result.

\begin{theorem}\label{theorem_ddsos}
  (\emph{Data-driven nonlinear stabilization using SOS})
  For the nonlinear polynomial system (\ref{dynamics_non poly simple}), under Assumptions \ref{assumption_equilibirium} and \ref{assumption_PE}, if there exists a matrix polynomial $Y(x)\in\R^{T\times N}$ such that
  \begin{itemize}
  \item[(i)] $\mathcal{Z}_{0,T}Y(x)=P$ where $P\succ 0$ is a symmetric constant matrix, and
  \item[(ii)] the matrix polynomial
  \begin{align*}
    Q(x)\!&:= -\Big[ \frac{\partial Z}{\partial x}X_{1,T}Y(x)
    \!+\!Y(x)^{\top}X_{1,T}^{\top}\Big( \frac{\partial Z}{\partial x} \Big)^{\top}\Big]\\
    & \quad~ -\! \epsilon(x)I_{N}
  \end{align*}
  is SOS for some SOS polynomial $\epsilon(x)$,
  \end{itemize}
  then the controller
  \begin{align}\label{controller_detailed}
  u= U_{0,1,T}Y(x)(\mathcal{Z}_{0,T}Y(x))^{-1}Z(x)
  \end{align}
  stabilizes the polynomial system. Moreover, if $\epsilon(x)> 0$ for all $x\ne 0$, the zero equilibrium is globally asymptotically stable.
\end{theorem}
\medskip

\proof
Consider the Lyapunov function candidate
\begin{align}
V(x)=Z(x)^{\top}P^{-1}Z(x)
\end{align}
with symmetric constant matrix $P\succ 0$ given in (i). The use of $P^{-1}$ instead of $P$ in the Lyapunov function is motivated by computational convenience. Taking directional derivative of $V(x)$ and using the definition of $Q(x)$ gives
\begin{align*}
  \dot V(x)
&= Z(x)^{\top} P^{-1}
\frac{\partial Z}{\partial x}  (A  + B  F(x)) Z(x)\\
&\quad\quad\quad\quad + Z(x)^{\top}(A  + B F(x))^{\top} \Big(\frac{\partial Z}{\partial x}\Big)^{\top} P^{-1} Z(x)\\
&= Z(x)^{\top}P^{-1} \Big[
\frac{\partial Z}{\partial x}  (A  + B  F(x)) P \\
&\quad\quad\quad\quad +P (A  + B  F(x))^{\top} \Big( \frac{\partial Z}{\partial x}\Big)^{\top}
\Big] P^{-1}  Z(x) 
\end{align*}
where $F(x)=U_{0,1,T}Y(x)(\mathcal{Z}_{0,T}Y(x))^{-1}$ by (\ref{controller_detailed}). Bearing in mind Lemma \ref{lemma_data rep}, we have
\begin{align}
  A+BF(x)=X_{1,T}G(x)
\end{align}
with $G(x)=Y(x)(\mathcal{Z}_{0,T}Y(x))^{-1}$. In fact, such a $G(x)$ satisfies (\ref{condition_Gx}) and $F(x)=U_{0,1,T}G(x)$.
As a result,
\begin{align*}
(A+BF(x))P &= X_{1,T}G(x)P\\
&=X_{1,T}\cdot Y(x)(\mathcal{Z}_{0,T}Y(x))^{-1}\cdot \mathcal{Z}_{0,T}Y(x)\\
&=X_{1,T} Y(x).
\end{align*}
Hence, we can express $\dot V(x)$ as
\begin{align*}
  \dot V(x)&=Z(x)^{\top}P^{-1}\Big[ \frac{\partial Z}{\partial x}X_{1,T}Y(x)\\
 &\quad\quad\quad\quad +Y(x)^{\top}X_{1,T}^{\top}\Big( \frac{\partial Z}{\partial x} \Big)^{\top} \Big]P^{-1}Z(x)\\
&= -Z(x)^{\top}P^{-1}(Q(x)+\epsilon(x)I_{N})P^{-1}Z(x).
\end{align*}
By Proposition \ref{prop_sos}, if $Q(x)$ is SOS, for all $x\in\R^{n}$
\begin{align*}
  Q(x)\succeq 0,
\end{align*}
which gives 
\begin{align*}
  \dot{V}(x)\le 0.
\end{align*}
Thus, the closed-loop system is stable at the equilibrium. Furthermore, if $\epsilon(x)>0$ for all $x\ne 0$, then $\dot{V}(x)< 0$ for all $x\in\R^{n}\setminus\{0\}$.  As $V(x)$ is radially unbounded, the closed-loop system is globally asymptotically stable at the zero equilibrium. The resulting stabilizing gain is
\begin{align*}
  F(x)= U_{0,1,T}G(x)= U_{0,1,T}Y(x)(\mathcal{Z}_{0,T}Y(x))^{-1}.
  \qed
\end{align*}

\subsection{Discussion}
\begin{enumerate}
\item \emph{On the choice of P}: In this work, the Lyapunov function is constructed with a constant matrix $P$ independent of $x$. In the data-driven stabilization case, this limits the choice of the Lyapunov functions, as the power vector $Z(x)$ should at least contain all monomials appearing in the nonlinear dynamics and thus cannot be chosen freely. It should be pointed out that, as discussed in \cite{Prajna2004SOS}, for the model-based stabilization, the success of Proposition \ref{prop_stability} depends on the choice of $Z(x)$. As we have observed in both model-based and model-free stabilizations, some choices of $Z(x)$ can make the SOS program infeasible. One way to overcome this issue is relaxing the choices of $Z(x)$ by making $P(x)$ a matrix polynomial in $x$. However, if $P(x)$ is dependent on $x$, the SOS program is no longer convex \cite{Prajna2004SOS}. Moreover, with $P(x)$, the Lyapunov function and controller can be rational functions, which makes the closed-loop analysis more complicated. In fact, by changing the data-based closed-loop representation, we can have more flexibility in the choices of the Lyapunov function and the control law so that the aforementioned issue can be overcome.
    \medskip
\item \emph{Computational complexity}: The key limitation of SOS optimization is its reduced scalability with respect to the dimension of the controlled system. For the stabilization of polynomial systems, the computational burden grows rapidly with the increase of either the dimension of the system, degree of the power vector $Z(x)$, or the size of $Z(x)$. In the model-based stabilization problems, variable $Y(x)$ is of size $m\times N$, where $m$ is the dimension of the input $u$. In our model-free learning control setting, the length of the experiment $T$ also affects the computational complexity. In particular, the size of the SOS program variable $Y(x)$ is $T\times N$ where $N$ is the size of $Z(x)$. To ensure that the data is informative enough for stabilization, Assumption \ref{assumption_PE} requires that $T\ge N$. Hence, the size of $Y(x)$ of the model-free case is mainly dependent on the size of $Z(x)$. In the subsequent section, the simulation example shows that if the size of $Z(x)$ is small, only a few data is required for the stabilizer design. Then, the computational complexity of data-driven stabilization is not substantially different from the model-free one. On the other hand, even if the size of $Z(x)$ is large, by representing the nonlinear system using a smaller vector, we can reduce the size of $Y(x)$ and effectively lighten the computational burden. This will be discussed in detail elsewhere.

    It is also noted that, to improve the scalability of SOS programming in general, \cite{Ahmadi2017scalabilitySOS} exploits the sparsity of the underlying SPD and \cite{Chesi2018complexitySOS} uses exact reduction methods to reduce the size of the LMIs and the number of LMI scalar variables. 
    \medskip

\item \emph{Noisy measurement of the derivative}: Because of the continuous-time nature of the nonlinear systems that we consider, the matrix $X_{1,T}$ contains the derivatives of the states at the sampling times. These measurements will be affected by noise. Following \cite[Section V.A]{DDC2019}, let us denote the noisy measurements as $Z_{1,T}=X_{1,T}+W_{1,T}$, where  $W_{1,T}$ is the unknown $n\times T$ matrix of noise vectors affecting the measurements. Then the {\em data-dependent polynomial}
representation of system \eqref{dynamics_non poly simple} changes into
$\dot x = (Z_{1,T} -W_{1,T})G (x) Z(x)$, which can be interpreted as a system with a nominal part,  $Z_{1,T} G (x) Z(x)$ and a perturbation, $-W_{1,T} G (x) Z(x)$, due to the effect of the noisy measurements. In \cite[Theorem 5]{DDC2019}, a robust stabilization result was given under a condition on  the signal-to-noise ratio. Other techniques from robust control can also used for the analysis of robustness  \cite{scherer-LMI-book,berberich2019data}. Techniques analogous to those in \cite{DDC2019} can be used to deal with noisy measurements in data-dependent polynomial representations $\dot x = (Z_{1,T} -W_{1,T})G (x) Z(x)$ .
\end{enumerate}

\section{Example}\label{sec_example}

Consider the nonlinear polynomial system
\begin{align*}
\dot{x}_{1} &= x_{2},\\
\dot{x}_{2} &= x_{1}^2+u,
\end{align*}
which is in the form of (\ref{dynamics_non poly simple}) with
\begin{align*}
A=\begin{bmatrix}
1 & 0 \\ 0 & 1
\end{bmatrix},~~
B=\begin{bmatrix}
0 \\ 1
\end{bmatrix},~~
Z(x)=\begin{bmatrix}
x_{2} \\ x_{1}^2
\end{bmatrix},
\end{align*}
$n=2$ and $N=2$.

\begin{figure}
  \centering
  \includegraphics[width=0.4\textwidth]{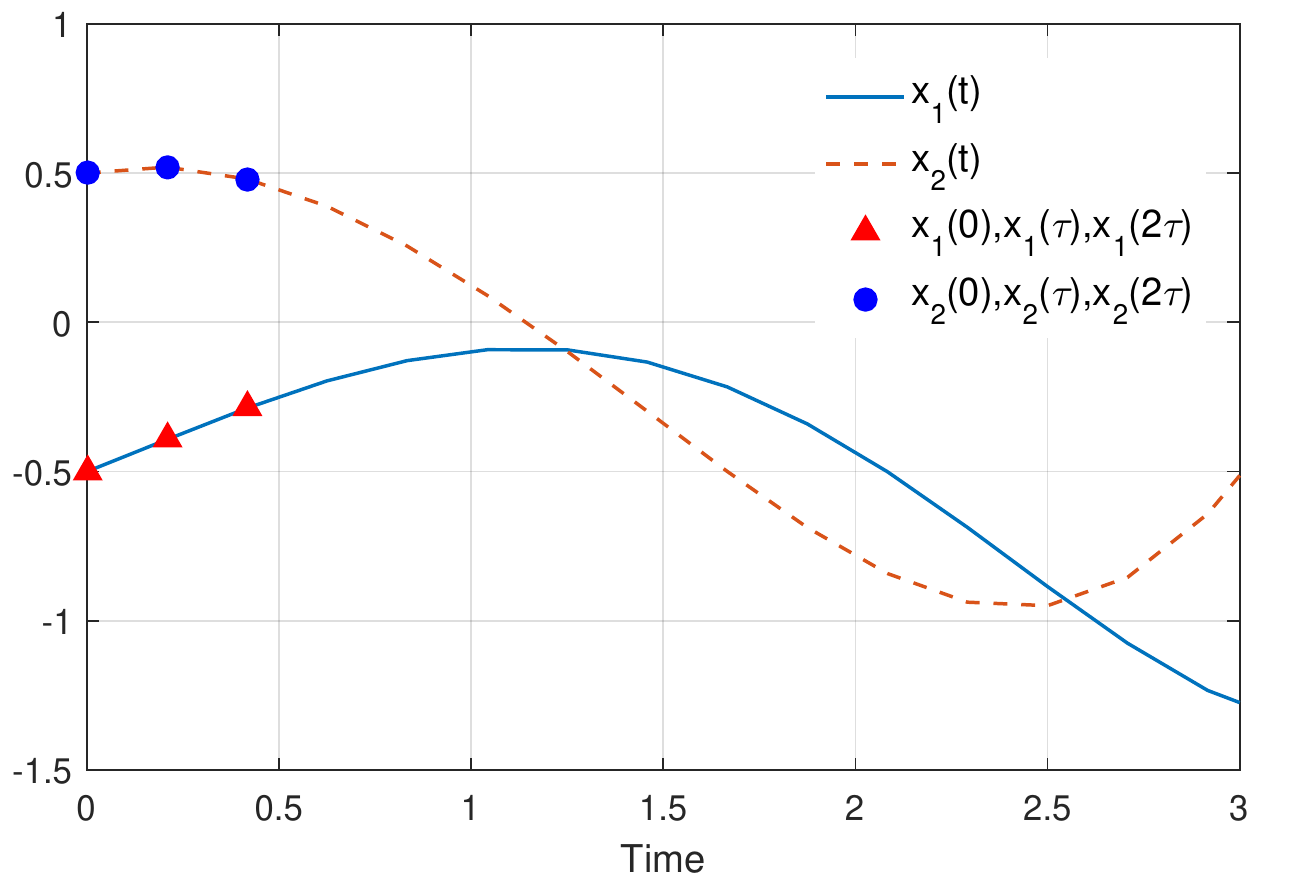}\\
  \caption{Experimental data (lines) and sampled output data (markers).}\label{Fig:experiment}
\end{figure}

An experiment is conducted with $u=-\sin(t)$ and $x(0)=[-0.5,0.5]$ during the time interval $t\in [0,3]$. The output data is sampled with $T=3$ as shown in Fig. \ref{Fig:experiment}. We arrange the data as
\begin{align*}
U_{0,1,T}&=\begin{bmatrix}
0 &	-0.2068 &	-0.4047
\end{bmatrix},\\
X_{0,T} & =\begin{bmatrix}
-0.5  & -0.3926  & -0.2874 \\
    0.5  &  0.5201  &  0.4804
\end{bmatrix},\\
X_{1,T} & =\begin{bmatrix}
0.5  &  0.5201  &  0.4804\\
    0.25 &  -0.0527 &  -0.3221
\end{bmatrix}.
\end{align*}
The data $\mathcal{Z}_{0,T}$ is calculated as
\begin{align*}
\mathcal{Z}_{0,T} & =\begin{bmatrix}
0.5  &  0.5201  &  0.4804 \\
    0.25 &   0.1542  &  0.0826
\end{bmatrix}.
\end{align*}

\begin{figure}
  \centering
  \includegraphics[width=0.4\textwidth]{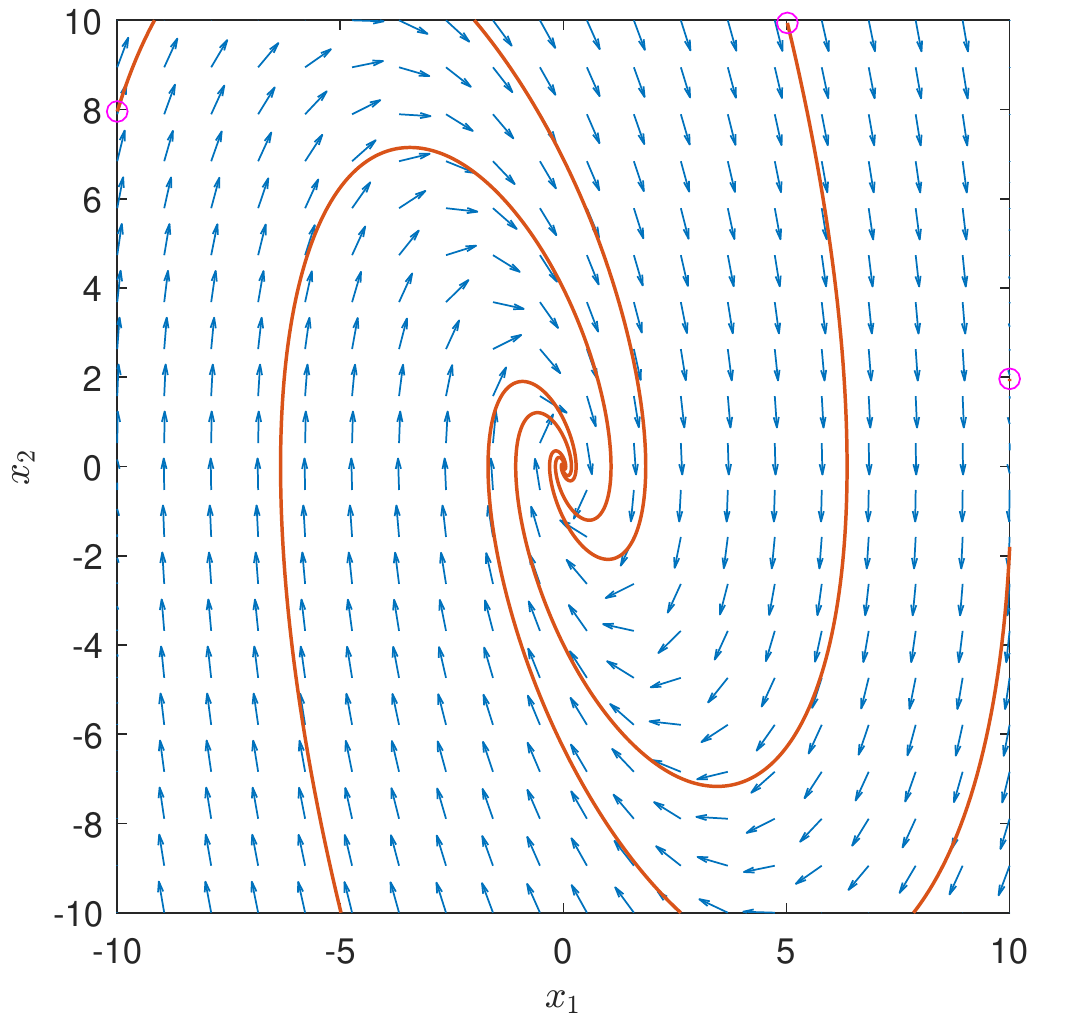}\\
  \caption{Phase portrait of the closed-loop system.}\label{Fig:closed-loop}
\end{figure}

Using the input-output data, we can formulate the SOS program to find a $3\times 2$ matrix polynomial $Y(x)$ in $x$ having degree $1$, such that $\mathcal{Z}_{0,T}Y(x)-\mu I_{N}$ is a constant SOS matrix for some $\mu>0$ and
\begin{align*}
&-\!\frac{\partial Z}{\partial x}X_{1,T}Y(x)
    \!-\!Y(x)^{\top}X_{1,T}^{\top}\Big( \frac{\partial Z}{\partial x} \Big)^{\top}\! -\! \epsilon(x)I_{N}
\end{align*}
is SOS for an SOS polynomial $\epsilon(x)$. Setting $\mu=10^{-3}$ and $\epsilon(x)=10^{-5}(x_1^2+x_{2}^2)$, the {\tt SOSTOOLS} gives the solution
\begin{align*}
Y(x)&=\begin{bmatrix}
 0.0224 &  0.0789 x_1 + 0.0858 \\
- 0.0741  &  - 0.2001x_1 - 0.1695 \\
  0.0705   &  0.1345x_1  + 0.0943 \\
\end{bmatrix}.
\end{align*}
Then, the constant matrix $P$ is
\begin{align*}
P=\begin{bmatrix}
  0.0065 &  0 \\
  0 &  0.0031
\end{bmatrix}
\end{align*}
and the feedback controller is
\begin{align*}
u=-2.0247x_{2}-4.2114x_{1}^3-x_{1}^2.
\end{align*}
The closed-loop system is globally asymptotically stable at the origin as illustrated in the phase portrait in Fig. \ref{Fig:closed-loop}.

\section{Conclusions and Future Work}\label{sec_conclusion}
Motivated by the fact that the identification of nonlinear systems is complex and time consuming, we proposed a control design method that learns a feedback control law directly from experimental data. In this work, we exploited the similarity between a class of nonlinear polynomial systems and linear systems, and constructed a stabilizing feedback controller using the SOS technique. The resulting SOS problem can be solved through SDPs and thus is computational tractable. The proposed approach has great potential in model-free learning control of more complex nonlinear systems. Some extensions of this work include seeking for data-based representations such that the choices on the Lyapunov functions and control laws can be more flexible. We may also consider searching for a more general class of Lyapunov function candidates using SOS program as shown in \cite{Majumdar2013ICRA}. Furthermore, future investigations may as well consider learning control of non-polynomial and/or uncertain nonlinear systems \cite{Chesi05domainof,Chesi09estimating,Hancock2013absolute&sos,Anderson2015AdvanceSOS}, nonlinear discrete-time systems \cite{Xu2007DTSOS,Valmorbida2012ACC}, and nonlinear optimal control using the SOS technique \cite{Papachristodoulou2002,Ichihara2009tac}.

%
%
%

\bibliographystyle{IEEEtran}
\bibliography{referenceDD}

\begin{thebibliography}{10}
\providecommand{\url}[1]{#1}
\csname url@samestyle\endcsname
\providecommand{\newblock}{\relax}
\providecommand{\bibinfo}[2]{#2}
\providecommand{\BIBentrySTDinterwordspacing}{\spaceskip=0pt\relax}
\providecommand{\BIBentryALTinterwordstretchfactor}{4}
\providecommand{\BIBentryALTinterwordspacing}{\spaceskip=\fontdimen2\font plus
\BIBentryALTinterwordstretchfactor\fontdimen3\font minus
  \fontdimen4\font\relax}
\providecommand{\BIBforeignlanguage}[2]{{%
\expandafter\ifx\csname l@#1\endcsname\relax
\typeout{** WARNING: IEEEtran.bst: No hyphenation pattern has been}%
\typeout{** loaded for the language `#1'. Using the pattern for}%
\typeout{** the default language instead.}%
\else
\language=\csname l@#1\endcsname
\fi
#2}}
\providecommand{\BIBdecl}{\relax}
\BIBdecl

\bibitem{Kerschen2006identification}
G.~Kerschen, K.~Worden, A.~F. Vakakis, and J.-C. Golinval, ``Past, present and
  future of nonlinear system identification in structural dynamics,''
  \emph{Mechanical Systems and Signal Processing}, vol.~20, no.~3, pp.
  505--592, 2006.

\bibitem{campi2006direct}
M.~Campi and S.~Savaresi, ``Direct nonlinear control design: {The} virtual
  reference feedback tuning ({VRFT}) approach,'' \emph{IEEE Transactions on
  Automatic Control}, vol.~51, no.~1, pp. 14--27, 2006.

\bibitem{Fliess2009}
M.~Fliess and C.~Join, ``Model-free control and intelligent {PID} controllers:
  {Towards} a possible trivialization of nonlinear control?'' \emph{IFAC
  Proceedings Volumes}, vol.~42, no.~10, pp. 1531--1550, 2009.

\bibitem{Tabuada2017CDC}
P.~Tabuada, W.-L. Ma, J.~Grizzle, and A.~D. Ames, ``Data-driven control for
  feedback linearizable single-input systems,'' in \emph{Proceedings of the
  56th Conference on Decision and Control (CDC)}, Melbourne, VIC, Australia,
  2017, pp. 6265--6270.

\bibitem{Lee2005nonlinearQlearning}
J.~M. Lee and J.~H. Lee, ``Approximate dynamic programming-based approaches for
  input-output data-driven control of nonlinear processes,'' \emph{Automatica},
  vol.~41, no.~7, pp. 1281--1288, 2005.

\bibitem{DDC2019}
C.~{De Persis} and P.~{Tesi}, ``Formulas for data-driven control:
  Stabilization, optimality, and robustness,'' \emph{IEEE Transactions on
  Automatic Control}, vol.~65, no.~3, pp. 909--924, 2020.

\bibitem{Prajna2004SOS}
S.~Prajna, A.~Papachristodoulou, and F.~Wu, ``Nonlinear control synthesis by
  sum of squares optimization: A {L}yapunov-based approach,'' in
  \emph{Proceedings of the 5th Asian Control Conference (ASCC)}, Melbourne,
  VIC, Australia, 2004, pp. 157--165.

\bibitem{Choi1995sos}
M.~D. Choi, T.~Y. Lam, and B.~Reznick, ``Sums of squares of real polynomials,''
  \emph{Proceedings of Symposia in Pure Mathematics}, vol.~58, no.~2, pp.
  103--126, 1995.

\bibitem{ParriloThesis}
P.~A. Parrilo, ``Structured semidefinite programs and semialgebraic geometry
  methods in robustness and optimization,'' Ph.D. dissertation, California
  Institute of Technology, Pasadena, California, 2000.

\bibitem{Chesi2009Book}
G.~Chesi, A.~Garulli, A.~Tesi, and A.~Vicino, \emph{Homogeneous Polynomial
  Forms for Robustness Analysis of Uncertain Systems}.\hskip 1em plus 0.5em
  minus 0.4em\relax New York: Springer, 2009.

\bibitem{Chesi2010}
G.~Chesi, ``{LMI} techniques for optimization over polynomials in control: {A}
  survey,'' \emph{IEEE Transactions on Automatic Control}, vol.~55, no.~11, pp.
  2500--2510, 2010.

\bibitem{tedrake2018iros}
W.~{Han} and R.~{Tedrake}, ``Controller synthesis for discrete-time polynomial
  systems via occupation measures,'' in \emph{Proceedings of the 2018 IEEE/RSJ
  International Conference on Intelligent Robots and Systems (IROS)}, Madrid,
  Spain, 2018, pp. 6911--6918.

\bibitem{Pirkelmann2019nolcos}
S.~Pirkelmann, D.~Angeli, and L.~Gr{\"u}ne, ``Approximate computation of
  storage functions for discrete-time systems using sum-of-squares
  techniques,'' in \emph{Proceedings of the 11th IFAC Symposium on Nonlinear
  Control Systems (NOLCOS)}, vol.~52, Vienna, Austria, 2019, pp. 508--513.

\bibitem{Ahmadi2017scalabilitySOS}
A.~A. Ahmadi, G.~Hall, A.~Papachristodoulou, J.~Saunderson, and Y.~Zheng,
  ``Improving efficiency and scalability of sum of squares optimization:
  {R}ecent advances and limitations,'' in \emph{Proceedings of the 56th IEEE
  Conference on Decision and Control (CDC)}, Melbourne, VIC, Australia, 2017,
  pp. 453--462.

\bibitem{Chesi2018complexitySOS}
G.~Chesi, ``On the complexity of {SOS} programming and applications in control
  systems,'' \emph{Asian Journal of Control}, vol.~20, no.~5, pp. 2005--2013,
  2018.

\bibitem{scherer-LMI-book}
C.~Scherer and S.~Weiland, ``Linear matrix inequalities in control,''
  \emph{Lecture Notes, Dutch Institute for Systems and Control}, 2019.

\bibitem{berberich2019data}
J.~Berberich, J.~K{\"o}hler, M.~M{\"u}ller, and F.~Allg{\"o}wer, ``Data-driven
  model predictive control with stability and robustness guarantees,''
  \emph{arXiv {\rm preprint} arXiv:1906.04679}, 2019.

\bibitem{Majumdar2013ICRA}
A.~Majumdar, A.~A. Ahmadi, , and R.~Tedrake, ``Control design along
  trajectories with sums of squares programming,'' in \emph{Proceedings of the
  2013 IEEE International Conference on Robotics and Automation (ICRA)},
  Karlsruhe, Germany, 2013, pp. 4054--4061.

\bibitem{Chesi05domainof}
G.~Chesi, ``Domain of attraction: {Estimates} for non-polynomial systems via
  {LMI}s,'' in \emph{Proceedings of the 16th IFAC World Congress on Automatic
  Control}, 2005.

\bibitem{Chesi09estimating}
------, ``Estimating the domain of attraction for non-polynomial systems via
  {LMI} optimizations,'' \emph{Automatica}, vol.~45, no.~6, pp. 1536--1541,
  2009.

\bibitem{Hancock2013absolute&sos}
E.~J. Hancock and A.~Papachristodoulou, ``Generalised absolute stability and
  sum of squares,'' \emph{Automatica}, vol.~49, no.~4, pp. 960--967, 2013.

\bibitem{Anderson2015AdvanceSOS}
J.~Anderson and A.~Papachristodoulou, ``Advances in computational {L}yapunov
  analysis using sum-of-squares programming,'' \emph{Discrete and Continuous
  Dynamical Systems Series B}, vol.~20, no.~8, pp. 2361--2381, 2015.

\bibitem{Xu2007DTSOS}
J.~Xu, L.~Xie, and Y.~Wang, ``Synthesis of discrete-time nonlinear systems: {A}
  {SOS} approach,'' in \emph{Proceedings of the 2007 American Control
  Conference (ACC)}, New York City, USA, 2007, pp. 4829--4834.

\bibitem{Valmorbida2012ACC}
G.~Valm{\'o}rbida, S.~Tarbouriech, G.~Garcia, and L.~Zaccarian, ``Synthesis of
  polynomial static state feedback laws and analysis for discrete-time
  polynomial systems with saturating inputs,'' in \emph{Proceedings of the 2012
  American Control Conference (ACC)}, Montreal, QC, 2012, pp. 2325--2330.

\bibitem{Papachristodoulou2002}
A.~Papachristodoulou and S.~Prajna, ``On the construction of {Lyapunov}
  functions using the sum of squares decomposition,'' in \emph{Proceedings of
  the 41st IEEE Conference on Decision and Control (CDC)}, Las Vegas, NV, USA,
  2002, pp. 3482--3487.

\bibitem{Ichihara2009tac}
H.~Ichihara, ``Optimal control for polynomial systems using matrix sum of
  squares relaxations,'' \emph{IEEE Transactions on Automatic Control},
  vol.~54, no.~5, pp. 1048--1053, 2009.

\end{thebibliography}

\end{document}